\documentstyle[aps,prd]{revtex}
\input epsf
\newcommand{\ri}{ | 1\rangle}
\newcommand{\ro}{| 0\rangle}
\newcommand{\li}{\langle 1 |}

\newcommand{\lpl}{\langle +|}
\newcommand{\rpl}{| +\rangle}

\begin{document}
\title{POVM optimization of classical correlations}
\maketitle
\begin{center}
S. Hamieh\\[0.3cm]
{\it Kernfysisch Versneller Instituut,\\
Zernikelaan
25
9747 AA Groningen,
The Netherlands.}
\end{center}
\vspace*{0.1cm}
\begin{center}
R. Kobes, H. Zaraket \\[0.3cm]
{\it Physics Dept, The University of Winnipeg,\\
515 Portage Avenue,
Winnipeg, Manitoba R3B 2E9, Canada.}
\end{center}

\begin{abstract}
We study the problem of optimization over positive valued-operator
measure to extract classical correlation in a bipartite quantum
system. The proposed method is applied to binary states only.
Moreover, to illustrate this method, an explicit example is
studied in details.
\end{abstract}
\section{Introduction}
Quantum computing constitutes a rich research area of physics and
computing at the same time. It is believed that, the expected
power of a quantum computer is derived from genuine quantum
resources. Entanglement, and {\it correlations} in general, are
typical quantum resources. However, not all correlations have pure
quantum nature. Generically, total correlations are ``mixture'' of
classical and quantum correlations. An important issue is to know
to what extent classical correlations are used in teleportation
protocols and quantum algorithms. For example, if one is able to
determine the classical part of correlations then by the optimal
measurement he can extract some information under in
classical form leaving the quantum state with less entropy. For this procedure
to be useful it should be done while retaining the ability to
regenerate the source state exactly from the classical measurement
result and the post-measurement state of the quantum system. This
has been studied in \cite{BenneBJMPSW1}. A possible application of
this is to send the post-measurement state through a noiseless
channel while sending the classical information through a more
robust channel.

Quantifying correlations implies measurement in most cases, which
is a non trivial task for quantum states. A general measurement
strategy is described by a positive operator-valued measure (POVM)
that decomposes the unity in the Hilbert space of the particle
under measurement. For the different measures of correlations (or
information) one has to find the optimal POVM, which can be
projective (orthogonal/nonorthogonal) or non-projective, and then
the number of elements in the optimal POVM. POVM optimization is
studied in different contexts in quantum information theory like
the accessible information, the cost function, Fidelity, .....
Davies' theorem \cite{Davie1} gives an upper bound for the maximum
number of elements of an optimal POVM needed to attain the {\bf
accessible information} (or the mutual information) of an ensemble of states in a
$d$-dimensional Hilbert space, this bound is $d^2$. The upper
bound has been reduced to $d(d+1)/2$ for states that are real
\cite{SasakBJOH1}. There were attempts to reduce this bound to $d$
elements, but Shor \cite{Shor1} has found an ensemble of states in
a three dimensional Hilbert space that was optimized with six
elements POVM. This eliminates the possibility of further
reduction of the upper bound. Similar POVM optimization was made
for the {\it fidelity} in \cite{FachsS1}, where for an ensemble of
quantum states in 2-dimensional Hilbert the optimal POVM was found
to have 3 elements, which fits the $d(d+1)/2$ bound. In the
present paper we present a POVM optimization for the classical
correlations. The study leads to two important results: for
arbitrary binary states projective POVM are found to be sufficient
to optimize classical correlation which resembles the cases
studied in \cite{Davie1,SasakBJOH1}, on the other hand the optimal
POVM is found to have at most 4 ($=2^2$) elements with indications
that the number can be reduced to 3 ($=2(2+1)/2$) for real states.
Moreover, for specific states the optimal POVM is shown to be
orthogonal. Another important property that is shown in the paper
is the possibility of decomposing the mutual information contained
in a given bipartite system into classical correlation and quantum
discord.

\vspace{5mm} The paper is organized as follows: the definitions of
POVM and classical correlation and their physical and mathematical
properties are given in the first two sections of the paper. Then
we show the natural decomposition of the mutual information into
classical correlation and quantum discord. The remaining part of
the paper is devoted to the complete optimization procedure,
including numerical simulation, of the classical correlation for a
particular state illustrating the need for the quantum discord in
order to match the mutual information.

\section{POVM}
Let ${\cal B}=\{B_i \}$ be a POVM, then $B_i$ should be a positive
valued operator, to preserve the positivity of the measurement
outcome probability. Each POVM ''set" should decompose the unity ,
to get probability completeness, {\it i.e.}
\begin{equation}
 \sum B_i=1\; .
\label{eq:complete}
\end{equation}
For later use it should be mentioned that the set of all POVM's is
convex. {\it i.e.} the {\it segment} joining two POVM's is a POVM.
In simpler terms: if ${\cal B}_1$ and ${\cal B}_2$ are two POVM's
and $0<\lambda<1$ (a probability) then
$$ \lambda{\cal B}_1+(1-\lambda){\cal B}_2$$
is indeed a POVM.

\section{Classical correlation}
Unfortunately, there is no unique measure of classical
correlations. Different measures are found in the literature
\cite{VedrPRK1,VedrP1,HendV1,OppenHHHH1,Ham03,DevetW1}. Two of the
most relevant requirements that should be satisfied by any
classical correlation are: (I) the classical correlations of
product states ($\rho_{AB}=\rho_A\otimes\rho_B$) should be zero.
(II) Classical correlation should not be affected by local unitary
transformations, which corresponds simply to a change of basis.
Other important conditions are listed in \cite{HendV1,OppenHHHH1}.
Comparison between different measures is found in the
literature. For example, property II has been investigated
recently by B. Synak and M. Horodecki \cite{SynakH1} for the
measure proposed in \cite{OppenHHHH1}, called the classical
information deficit ($\Delta_{cl}^{\rightarrow}$). It is found
that $\Delta_{cl}^{\rightarrow}$ does increase under local
operations. Moreover it was shown \cite{SynakH1} that
$\Delta_{cl}^{\rightarrow}$ is bounded above by the classical
correlation measure proposed in \cite{HendV1}. The only case where
$\Delta_{cl}^{\rightarrow}$ is monotone under local operation is
when it coincides with the classical correlation of \cite{HendV1}.
This comparison can be considered to be in favor of the
Henderson and Vedral measure. However there is no evidence that
the other measures of classical correlations violate some of
the expected physical properties of a correlation. So one can
choose {\bf a} classical correlation measure out of the proposed measures.
Our choice is the Henderson and Vedral measure.

Given the bipartite state  $\rho_{AB}$ a possible measure of the
classical correlation between subsystems $A$ and $B$ is
(\cite{HendV1})
\begin{equation}
C_{B}(\rho_{AB})=\max_{\cal
B}\left[S(\rho_A)-\sum_{i}p_iS(\rho_A^i)\right]\; ,
\end{equation}
where $\rho_A={\rm tr}_B(\rho_{AB})$ is the reduced density
matrix. The Von Neumann entropy is $S(\rho)=-{\rm
tr}(\rho\log\rho)$. ${\cal B}$ is a POVM and the sum over $i$ runs
over all its elements $B_i$. The conditional density matrix
$\rho_A^i$ is the density matrix of $A$ after performing the
measurement $B_i$ on $B$:
\begin{equation}
\rho_A^i=\frac{{\rm tr}_B\left(B_i\rho_{AB}\right)}{{\rm
tr}_{AB}\left(B_i\rho_{AB}\right)}\; \; .
\end{equation}
The probability of $A$ being in the state $\rho_A^i$ is $p_i={\rm
tr}_{AB}(B_i\rho_{AB})$.

The correlation measure $C_B$ has a simple physical
interpretation: if $A$ and $B$ are not correlated then the
marginal entropy of $A$ ($S(\rho_A)$) and the residual entropy of
$A$ after a POVM measurement on $B$ ($\sum p_iS(\rho_A^i)$) should
coincide to give $C_B=0$, since for uncorrelated system $AB$, $A$
is not affected by a POVM measurement on $B$. Moreover, note that
$$\rho_A= \sum_{i}p_i\rho_A^i$$
hence for a {\bf given} POVM ${\cal B}$ the combination
$$ S(\rho_A)-\sum_{i}p_iS(\rho_A^i)$$
is closely related to the {\it entropy defect} defined by Levitin
(see \cite{Levit1} and references therein). So the classical
correlation $C_B$ can be seen as the maximum average decrease in the
entropy of the system $A$ when a state $\rho_A^i$ (after a measurement
$B_i$ is preformed on $B$) is specified compared with the situation
when only the mixture of states $\rho_A$ is known.

The correlation $C_A(\rho_{AB})$ is obtained from $C_B$ by making
the replacement $A\leftrightarrow B$ in the above formulas. It is
evident that the measure $C_{A,B}$ are not manifestly symmetric
under the exchange of the roles of $A$ and $B$. Whereas, one would
expect that the classical correlation is a measure of how strongly
the two subsystems are correlated no matter which subsystem is used to
extract such a correlation.
Hence, until a formal proof of an explicit symmetry of this
measure, if it exists, the above proposed measure can not be
considered as a universal measure of the existing classical
correlations.

\section{POVM choice and convexity}
In what follows we consider binary states exclusively. Extension
to higher dimension is not straight forward.
\subsection{Convexity}
\label{sec:concavity} An important property that helps in reducing
the range of exploration in the set of all POVM's is the
concavity \cite{Rbhat1} of the classical correlation measure. If
one is able to show that the classical correlation is convex then
(see \cite{Davie1}), for binary states, one can consider POVM's
of rank one elements that can be taken to be proportional to the
one-dimensional projectors.

{\bf Proof:}

For binary states the Von Neumann entropy is concave {\it i.e.}
\begin{equation}
S(\lambda\rho_1+(1-\lambda)\rho_2)\geq \lambda
S(\rho_1)+(1-\lambda)S(\rho_2)
\end{equation}
for all $0<\lambda<1$.

To study the concavity of the classical measure it suffices to
study the variation of the functional $F(\rho_A,{\cal B})$ defined
as
\begin{equation}
F(\rho_A,{\cal B})=-p_{\cal B}S(\rho_A^{\cal B})\; ,
\end{equation}
where the compact notation: $p_{\cal B}={\rm tr}_{AB}\left({\cal
B}\rho_{AB}\right)$ and $\rho_A^{\cal B}={\rm tr}_B\left({\cal
B}\rho_{AB}\right)/p_{\cal B}$ is used for all elements $B_i$ of
the POVM ${\cal B}$.

Let ${\cal C}$ and ${\cal D}$ be two POVM's and let ${\cal G}$ be
their combination
\begin{equation}
{\cal G}=\lambda{\cal C}+(1-\lambda){\cal D}\; ,
\end{equation}
for $0<\lambda<1$. ${\cal G}$ is a POVM since, as mentioned
before, the set of all POVM's is convex.

It is straight forward to show that
\begin{eqnarray}
p_A^{\cal G}&=&\lambda p_A^{\cal C}+(1-\lambda)p_A^{\cal D}\\
\rho_A^{\cal G}&=&\lambda \frac{p_A^{\cal C}}{p_A^{\cal
G}}\rho_A^{\cal C}+(1-\lambda) \frac{p_A^{\cal D}}{p_A^{\cal
G}}\rho_A^{\cal D}\; .
\end{eqnarray}
Using these equations and the fact that the entropy is a concave
function, it can be shown that
\begin{equation}
F(\rho_A,{\cal G})\leq \lambda F(\rho_A,{\cal C})
+(1-\lambda)F(\rho_A,{\cal D}).
\end{equation}
Therefore, $F$ is a convex over the convex set of all POVM's.
Hence the maximum of $F$ occurs for an {\it extremal} POVM \cite{Trock1}. For
binary states it was explicitly shown recently \cite{ArianPP1} that extreme
POVM's are of rank one\footnote{For higher
dimensional Hilbert spaces there are higher rank extreme POVM's
\cite{ArianPP1}.}. Hence, to optimize $F$ and consequently the
classical correlation, since $S(\rho_A)$ is independent of the
POVM performed, one can consider the special class of POVM with
rank one elements. Hence we are left with projective measurements
only\footnote{It should be mentioned that for generic cases one
can use Naimark theorem to get projective POVM only without being
restrictive, but this is done at the cost of extending the
original Hilbert space.}.

An important consequence of this result is presented in the next
section.
\subsection{Quantum discord and mutual information}
As mentioned in \cite{SynakH1} the quantum mutual information
$I(A:B)$ of a bipartite system $AB$ can be decomposed into an
information deficit (or work deficit) $\Delta$ and the classical
information deficit $\Delta_{cl}$\footnote{The previously
mentioned classical information deficit
$\Delta_{cl}^{\rightarrow}$ is a one-way quantity {\it i.e.} for
a system with two parties $A$ and $B$ we allow communication from
$A$ to $B$ only or vice versa. This asymmetric definition is
analogous to the seemingly asymmetric property of the classical
correlations where one has to perform measurements on one of the
two parties.}:
\begin{equation}
I=\Delta_{cl}+\Delta\; .
\end{equation}
So, naturally one should be concerned with an analogous
decomposition of the mutual information when using a different
measure of classical correlation.

By analyzing different quantum states, the authors of
\cite{HendV1} found that the estimated classical correlation and
the relative entropy of entanglement ($E_{RE}$) do not add up to
give the Von Neumann mutual information between the two
subsystems, {\it i.e.}
\begin{equation}
\left(C_B(\rho_{AB})\right)_{\rm optimized} +E_{RE}<
I(\rho_{A:B})\; .
\end{equation}
It was argued that either the mutual information is not the best
quantity to measure the total correlations, or that a more
elaborated choice of the POVM may saturate the total correlations.
By considering all possible POVM for a given state, we show that
the most optimal POVM can not saturate the total correlations.
This implies that: either the mutual information is not a good
measure of total correlations, which is highly improbable, or one
has to change the measure $C_B$ of classical correlation. The
third alternative, which renders the definition of classical
correlations compatible with the mutual information, is the
possibility of having a different definition of quantum
correlations which is different from $E_{RE}$. A possible
candidate is the quantum discord, which was first defined in
\cite{OllivZ1}. It is the result of the difference between
classical and quantum conditional entropies. In contrast to
classical conditional entropy, quantum conditional entropy is a
measurement dependent quantity (see further \cite{LevitT1}). The
quantum discord is defined as \cite{OllivZ1}
\begin{equation}
\delta(A:B)=I(A:B)-J(A:B)_{\{\Pi_i^B\}}\; .
\end{equation}
$J$ is the information gained about $B$ as a result of the
set of measurements $\{\Pi_i^B\}$:
\begin{equation}
J(A:B)_{\{\Pi_i^B\}}=S(A)-S(A|\{\Pi_i^B\})=S(A)-\sum
{}^*p_iS(^*\rho_A^i)\; .
\end{equation}
where ${}^*p_i$ and ${}^*\rho_A^i$ were defined originally as the
special case of $p_i$ and $\rho_A^i$ when the choice of the set of
measurement is restricted to one dimensional projectors $\Pi_i^B$.
The above definition of the quantum discord reflects clearly the
inherited measurement dependence of the quantum conditional
entropy. For the quantum discord to be zero one has to find at
least one measurement for which it is zero. Therefore the minimum
of the quantum discord is the relevant quantum correlation. By
focusing on perfect measurements of $B$, defined by a set of one
dimensional projectors, one can easily check that the set of
measurements that minimizes the quantum discord, {\it i.e.}
maximizes $J$, is exactly the same POVM set that optimizes the
classical correlations {\it for binary states}. This follows from
the definition of both quantities, and the result of the previous
section on the optimization of classical correlation using
projective measurements only. Hence
\begin{equation}
{\rm Max}(J(A:B)_{\{\Pi_i^B\}})=C_B(\rho_{AB})\; .
\end{equation}
Therefore
\begin{equation}
I(A:B)=C_B+\min_{\{\Pi_i^B\}}\delta(A:B)\;,
\end{equation}
{\it i.e.} {\it for binary states the classical correlation and
the quantum discord add up to give the mutual information.}

\subsection{What is next?}
Limiting the domain of exploration in the set of all POVM's to
projective POVM leads to important simplification. But the game is
not over. What would be the optimal projective POVM? How many
elements there are in the optimal POVM? Corollary 1 of
\cite{ArianPP1} implies that POVM's with 5 or more elements acting on a two dimensional
Hilbert space are not extreme, and hence
cannot optimize the classical correlation\footnote{This is
analogous to the $d^2$ bound found in \cite{Davie1}.}. So, we are
left with POVM's having $2,3$ or four elements. In the next
section we illustrate, through an example, the steps that can be
followed to complete the optimization procedure.

\section{An example}

Consider a system $AB$ in a state $\rho_{AB}$ with probability $p$ to be
in a state $\rho_1$ and $(1-p)$ to be in another state $\rho_2$. A possible example is
the following state studied in \cite{HendV1}
\begin{equation}
p|0\rangle\langle 0|\otimes|0\rangle\langle 0| +
(1-p)\rpl\lpl\otimes\rpl\lpl\; , \label{eq:Vederal}
\end{equation}
where $\rpl=(\ro+\ri)/\sqrt{2}$. To enlarge the "library" of
computed classical correlations we will not use the above state
but rather a slightly modified version of it, namely
\begin{equation}
\rho_{AB}=p\ri \li \otimes \ri \li +
(1-p)\rpl\lpl\otimes\rpl\lpl\; . \label{eq:our}
\end{equation}
The classical correlation of the state in eq. (\ref{eq:Vederal})
was calculated in \cite{HendV1} using optimization over all {\it
orthogonal} measurements, which was not a justified restriction.

For the state given in eq. (\ref{eq:our}), it is useful to use the
Pauli matrices decomposition. Hence the measurement operator $B_i$
can be written as:
\begin{equation}
B_i=|b_i|\left(\frac{1+\vec{\sigma}\cdot{\hat b}_i}{2}\right)\; .
\end{equation}
The projection direction is set by the unit vector ${\hat b}_i$.
We should further require the completeness relation
(eq.~(\ref{eq:complete})):
\begin{equation}\label{eq:constraint}
\sum {\vec b}_i=0\; , \quad \sum |b_i|=2\; .
\end{equation}
The state $\rho_{AB}$ in eq. (\ref{eq:our}) becomes
\begin{equation}\label{eq:rhosigma}
\rho_{AB}=\frac{p}{2}\ri\li\otimes
(1-\sigma_z)+\frac{1-p}{2}\rpl\lpl\otimes (1+\sigma_x)\; .
\end{equation}
It is straight forward to show that the density $\rho_A^i$ of subsystem $A$,
after the measurement $B_i$ on $B$, is
\begin{equation}
\rho_A^i=\alpha_i\ri\li+(1-\alpha_i)\rpl\lpl
\end{equation}
where
\begin{equation}
\alpha_i=\frac{p(1-{\hat b}_i\cdot {\hat e}_z)}{p(1-{\hat b}_i\cdot {\hat e}_z)+(1-p)(1+{\hat b}_i\cdot {\hat e}_x)}.
\end{equation}
Hence the entropy of the subsystem $A$ after the measurement $B_i$ of the subsystem $B$ is
\begin{equation}
S(\rho_A^i)=h(\omega_i^{(1)})
\end{equation}
where as usual $h(x)=-x\log(x)-(1-x)\log(1-x)$ is the Shannon's entropy function
and
\begin{equation}
\omega_i^{(1),(2)}=\frac{1}{2}(1\pm\sqrt{\alpha_i^2+(1-\alpha_i)^2})\;
.
\end{equation}
are the eigenvalues of $\rho_A^i$.

The probability $p_i$ to have the subsystem $A$ in the state $\rho_A^i$ is
\begin{equation}
p_i={\rm tr}_{AB}\left ( B_i\rho_{AB}\right )=\frac{|b_i|}{2}\left(p(1-{\hat b}_i\cdot {\hat e}_z)+(1-p)(1+{\hat b}_i\cdot {\hat e}_x)\right)\; .
\end{equation}

Note that, the same procedure can be applied for the state given
in eq. (\ref{eq:Vederal}). One has to replace $\sigma_z$ by
$-\sigma_z$ in eq.~(\ref{eq:rhosigma}). This change of sign will
propagate to all the subsequent equations.

\subsection{Choice of basis}
The previous section seems to underline a choice of the basis for
projection. One can argue that the component $({\hat b}_i)_y$ does
not show up in the conditional density matrix $\rho_A^i$, hence we
can set it to zero in the projection operator. In other words
measurement can be performed in the $xz$ plane. In general, this
is not correct it was shown in \cite{ArianPP1} (see section III)
that an extreme POVM with four outcomes ($n=4$) cannot have its
four vectors ${\hat b}_i$ coplanar. On the other hand, it is easy
to show that the entropy
$$ \sum p_i S(\rho_A^i)$$
is maximum, {\it i.e.} it is equal to the entropy of the reduced density matrix $\rho_A$
$$ \sum p_i S(\rho_A^i)=S(\rho_A)$$
for a projective measurement in the $y$ direction, since the state
of $B$ is completely random in that direction. However for
classical correlations the minimum of $\sum p_i S(\rho_A^i)$ is
needed not its maximum. Therefore, it is justified to use
measurements such that $({\hat b}_i)_y=0$. Hence $({\hat
b_i})_z=\cos\theta_i$ and $({\hat b_i})_x=\sin\theta_i$. But, with
this assumption the $n=4$ extreme POVM is excluded. This should be
investigated further to see whether it is a general property of
real states, to which the present state belongs, {\it i.e.}
whether optimal POVM's for classical correlation of real binary
states have at most 3 outcomes in analogy with the $d(d+1)/2$
bound found in \cite{SasakBJOH1}. In fact, the proof of
\cite{SasakBJOH1} (see Lemma 5 and theorem 1) relies on the
convexity of the accessible information and some of it scaling
properties under group transformation this seems to be equally
valid for the classical correlation. This will be studied in a
future work. But until such a rigorous proof is given the present
"exclusion" of the $n=4$ case should be considered as a reasonable
assumption.

\subsection{Optimization}

\subsubsection{ Method 1: Lagrange multipliers} The determination
of the classical correlation $C_B$ requires finding a POVM that
minimizes $\sum p_i S(\rho_A^i)$, given the constraint equations
(\ref{eq:constraint}). Hence a priori we have to explore all
possible POVM's and then picking up the one that minimizes the
considered sum. The result of section \ref{sec:concavity} has
limited the domain of exploration to projective POVM only. The
optimization problem can be done using the Lagrange multipliers
method. It is more convenient to recast the problem as
\begin{eqnarray}
{\rm min}\quad && \sum p_i S(\rho_A^i)= \sum r_if(\theta_i)\nonumber \\
{\rm subject\; to}\quad && \sum r_i -2=0\nonumber\\
&& \sum r_i\cos\theta_i=0\nonumber\\
&& \sum r_i\sin\theta_i=0\; , \label{1}
\end{eqnarray}
where $r_i=|b_i|$ and
$$f(\theta_i)= \frac{1}{2}\left[p(1-\cos\theta_i)+(1-p)(1+\sin\theta_i\right)]h(\omega_i^{(1)}(\theta_i))\; . $$
Let $\lambda_{1,2,3}$ be the Lagrange multipliers associated with the above
constraints. Hence the minimization problem is equivalent
to find the values of $\theta_i, r_i, \lambda_{1,2,3}$ satisfying
\begin{eqnarray}
f(\theta_i)+\lambda_1+\lambda_2\cos\theta_i+\lambda_3\sin\theta_i=0\\
r_i(\frac{df(\theta_i)}{d\theta_i}-\lambda_2\sin\theta_i+\lambda_3\cos\theta_i)=0\\
\sum r_i=2 \\
\sum r_i\cos\theta_i=0 \\
\sum r_i\sin\theta_i=0\; ,
\end{eqnarray}
This optimization should be done for $n=2,3$.

\subsubsection{Orthogonal projective measurements}
For two-elements POVM ($n=2$) we recover the orthogonal projective
measurement\footnote{If one takes $b_1=b_2=1$, it is straight
forward to check that the optimal POVM measurement is indeed
projective, {\it i.e.} $B_1=|\psi_1\rangle\langle\psi_1|$ and
$B_2=|\psi_2\rangle\langle\psi_2|$, with the bases of measurement
parameterized by $\theta$ as
$$ \{\cos(\frac{\theta}{2})\ro+\sin(\frac{\theta}{2})\ri,\sin(\frac{\theta}{2})\ro-\cos(\frac{\theta}{2})\ri\;.$$}, since
$\vec{b}_1+\vec{b}_2=0$. Hence $\theta_2= \theta_1+\pi$. The {\it
best} measurement, {\it i.e.} the optimal measurement among the
$n=2$ POVM's only, is obtained when
$f(\theta_1)=f(\theta_2)=f(\theta_1+\pi)$. Therefore, the {\it
best} set of measurement is a two-shot measurements with opposite
directions. This should be solved numerically for each $p$. The
result is plotted in figure \ref{fig:correl}. Our result matches
well with the result of \cite{HendV1}. This implies that although
our state is slightly different from the state they have studied
but the two states have the same classical correlations. However,
as mentioned before, this is just the orthogonal measurement, non
orthogonal measurements should be considered which was not done in
\cite{HendV1}.

\subsubsection{Method 2: Monte Carlo simulation $n=3$} Finding the
minimum by the method presented above (method 1) for $n=3$ seems
to be a non trivial task, however, it is possible to use a
different method like the Monte Carlo Simulations (MCS) method. We
generate random events that correspond to the set of the 3 angles
$\{\theta\}_{i=1}^{3}$, while from eq. \ref{1} the three $r_i$'s
are no longer independent variables. Thus our strategy is to
compute the classical correlations for each randomly generated
event. To each event a numerical value is obtained for the
classical correlation, for a given $p$, this is is represented by
the large number of points (forming long continuous strips) in
Fig. \ref{fig:correl}. In Fig. \ref{fig:correl} we plot the
simulation outputs for $10^6$ events. Clearly the classical
correlation predicted for the {\it best} orthogonal projective
measurement case constitutes the upper limit of all the results
obtained for the random events. Hence {\bf the} optimal POVM,
among all possible POVM's, is the two-outcomes orthogonal
projective POVM. This result can be linked to the remark found in
the conclusion of \cite{Levit1} that one should dare conjecture the following:
orthogonal measurement is optimal whenever the number of states
(the two states $\ri \li \otimes \ri \li$ and $\rpl\lpl\otimes\rpl\lpl$ in the present case)
does not exceed the dimensionality of the state space.

\subsubsection{Mutual information}
\begin{figure}[h]
\centerline{
    \mbox{\epsfxsize=10cm\epsffile{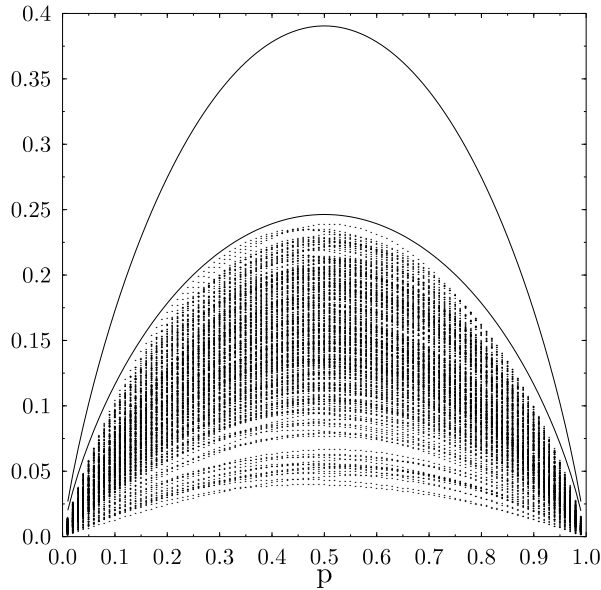}}
    }
\caption{\label{fig:correl} We plot as a function of
$p$: the classical correlation $C_B$ (lower curve) evaluated using
the {\it best} orthogonal POVM, and the mutual information
$I(A:B)$ (higher curve). The vertical strips are sets of large
numbers of dots representing the obtained value of the classical
correlation for the different randomly generated events.}
\end{figure}

To understand the role of the classical correlations and its
compatibility with the mutual information, the mutual information
is evaluated using the formula
\begin{equation}
I(A:B)=S(\rho_A)+S(\rho_B)-S(\rho_{AB})\; .
\end{equation}
For the present state of the system $AB$ the mutual information can be easily
calculated to obtain:
\begin{equation}
I(A:B)=2h(\frac{1+\sqrt{p^2+(1-p)^2}}{2})-h(\frac{1+\sqrt{1+3p^2-3p}}{2})\;
.
\end{equation}
$I(A:B$) is plotted in Fig. \ref{fig:correl} besides the classical
correlation obtained by the optimal orthogonal projective
measurement and the one generated by MCS. Since $\rho_{AB}$ is
separable, then its relative entropy of entanglement is zero.
Therefore, from our general POVM analysis, and as shown in Fig.
\ref{fig:correl}, no POVM can lead to classical correlation that
saturates, when added to the relative entropy of entanglement, the
mutual information between the two subsystems. This confirms the
necessity to use the quantum discord as a quantum counter part for
the classical correlation used in this paper rather than the
relative entropy of entanglement.

\section{Conclusions}
By considering a generic POVM we show that classical correlations
of binary states are optimized via projective POVM. It is found
that the classical correlation and the quantum discord add up to
give the mutual quantum information.

This work should be considered as a first step towards a
generalization of the POVM optimization for states in higher
dimensions and to see whether rank one POVM elements are still the
optimal choice for classical correlation. Moreover, there is at
present strong evidence on the possibility of an experimental
realization of non projective POVM using optical devices as was
recently proposed in \cite{AhrenP1}. This gives a new dimension to
POVM optimization.

This work was performed as part of the research program of the {\sl Stichting voor Fundamenteel Onderzoek
der Materie (FOM)} with financial support from the {\sl Nederlandse Organisatie voor Wetenschappelijk Onderzoek }.



\begin{thebibliography}{99}
\bibitem{BenneBJMPSW1}
C. H. Bennett, G. Brassard, R. Jozsa, D. Mayers, A. Peres, B.
Schumacher and W. K. Wootters, {\it J. Mod. Optics} {\bf 41}, 2307
(1994).
\bibitem{Davie1}
E. B. Davies, {\it IEEE Trans. Inf. Theory} {\bf IT-24}, 596
(1978).
\bibitem{SasakBJOH1}
M. Sasaki, S.M. Barnett, R. Jozsa, M. Osaki and O. Hirota, {\it
Phys. Rev. A} {\bf 59}, 3325 (1999).
\bibitem{Shor1}
P. W. Shor, {\it Proceeding of QCM\&C 2000}, Kluwer (2000).
\bibitem{FachsS1}
C.A. Fachus and M. Sasaki, quant-ph/0302092 (2003).
\bibitem{VedrPRK1}
V. Vedral, M. B. Plenio, M. A. Rippin and P. L. Knight,
  {\it Phys. Rev. Lett.} {\bf 78}, 2275 (1997),
quant-ph/9702027.
\bibitem{VedrP1}
V. Vedral and M. B. Plenio, {\it Phys. Rev. A}
{\bf 57}, 1619 (1998), quant-ph/9707035.
\bibitem{HendV1}
 L. Henderson and V. Vedral,
{\it J. Phys. A.  Math. \& Gen.} {\bf 34}, 6899  (2001).
\bibitem{OppenHHHH1}
 J. Oppenheim, K. Horodecki, M. Horodecki, P. Horodecki and R.
 Horodecki, {\it Phys. Rev. A} {\bf 68}, 022307 (2003).
\bibitem{Ham03} S. Hamieh, J. Qi, D. Siminovitch, and M. K. Ali,
   {\it Phys. Rev. A} {\bf 67},  014301 (2003).
 \bibitem{DevetW1}
  I. Devetak, A. Winter, quant-ph/0304196 (2003).
\bibitem{SynakH1}
 B. Synak, M. Horodecki, quant-ph/0403167 (2004).
 \bibitem{Levit1}
 L.B. Levitin, in {\it Quantum communications and Measurement}, Ed. V.P. Belavkin {\it et al}, pages 439-448,
 Plenum Presss, New York (1995).
\bibitem{Rbhat1}
For the properties of convex functions defined over operators in a
Hilbert space one can see: R. Bhatia, {\it Matrix analysis},
Springer-Verlag, New York (1997).
\bibitem{Trock1}
The supremum of a convex function defined over a closed convex set is
attained at some extreme points of the set provided that the set does not contain
any lines, which is the case for the set of all POVM's. See for example
Theorem 32.3 and Corollary 32.3.1 in: R.T. Rockafellar, {\it Convex analysis},
Princeton, New Jersey, Princeton University Press (1970).
\bibitem{ArianPP1}
G.M. D'Ariano, P.L Presti and P. Perinotti, quant-ph/0408115
(2004).
\bibitem{OllivZ1}
H. Ollivier and W. Zurek, {\it Phys. Rev. Lett.} {\bf 88}, 017901
(2002).
\bibitem{LevitT1}
L.B. Levitin and T. Toffoli, quant-ph/0306058 (2003).
\bibitem{AhrenP1}
S.E. Ahrent and M.C. Payne, quant-ph/0408011 (2004).
\end{thebibliography}
\end{document}